# The third five-parametric hypergeometric quantum-mechanical potential

T.A. Ishkhanyan[1,2] and A.M. Ishkhanyan[1,3]

[1]Institute for Physical Research, NAS of Armenia, Ashtarak, 0203 Armenia
[2]Moscow Institute of Physics and Technology, Dolgoprudny, 141700 Russian Federation
[3]Institute of Physics and Technology, National Research Tomsk Polytechnic University, Tomsk, 634050 Russian Federation

We introduce the third *five-parametric* ordinary hypergeometric energy-independent quantum-mechanical potential, after the Eckart and Pöschl-Teller potentials, which is proportional to an arbitrary variable parameter and has a shape that is independent of that parameter. Depending on an involved parameter, the potential presents either a short-range singular well (which behaves as inverse square root at the origin and vanishes exponentially at infinity) or a smooth asymmetric step-barrier (with variable height and steepness). The general solution of the Schrödinger equation for this potential, which is a member of a general Heun family of potentials, is written through fundamental solutions each of which presents an irreducible linear combination of two Gauss ordinary hypergeometric functions.

## 1. Introduction

The solutions of the Schrödinger equation in terms of special mathematical functions for energy-independent potentials which are proportional to an arbitrary variable parameter and have a shape independent of that parameter are very rare [1-10] (see the discussion in [11]). It is a common convention to refer to such potentials as *exactly* solvable in order to distinguish them from the *conditionally* integrable ones for which a condition is imposed on the potential parameters such that the shape of the potential is not independent of the potential strength (e.g., a parameter is fixed to a constant or different term-strengths are not varied independently). While there is a relatively large set of potentials of the latter type (see, e.g., [12-20] for some examples discussed in the past, and [21-25] for some recent examples), the list of the known exactly integrable potentials is rather limited even for the potentials of the most flexible *hypergeometric* class. The list of the exactly solvable hypergeometric potentials currently involves only ten items [1-10]. Six of these potentials are solved in terms of the confluent hypergeometric functions [1-6]. These are the classical Coulomb [1], harmonic oscillator [2] and Morse [3] potentials and the three recently derived potentials, which are the inverse square root [4], the Lambert-W step [5] and Lambert-W singular [6] potentials. The remaining four exactly integrable potentials which are solved in terms of the Gauss ordinary hypergeometric functions are the classical Eckart [7] and Pöschl-Teller [8] potentials and the two new potentials that we have introduced recently [9,10].

An observation worth to be mentioned here is that all five classical hypergeometric potentials, both confluent and ordinary, involve *five* arbitrary variable parameters, while all new potentials are four-parametric. In this communication we show that the two four-parametric ordinary hypergeometric potentials [9,10] are in fact particular cases of a more general five-parametric potential which is solved in terms of the hypergeometric functions. This generalization thus suggests the third five-parametric ordinary hypergeometric quantum-mechanical potential after the ones by Eckart [7] and Pöschl-Teller [8].

The potential we introduce belongs to one of the eleven independent eight-parametric general Heun families [25] (see also [26]). From the mathematical point of view, a peculiarity of the potential is that this is the only known case when the location of a singularity of the equation to which the Schrödinger equation is reduced is not fixed to a particular point but stands for a variable potential-parameter. Precisely, in our case the third finite singularity of the Heun equation, that located at a point $z=a$ of the complex $z$-plane (that is the singularity which is additional if compared with the ordinary hypergeometric equation), is not fixed but is variable - it stands for the fifth free parameter of the potential.

The potential is in general defined parametrically as a pair of functions $V(z), x(z)$. However, in several cases the coordinate transformation $x(z)$ is inverted thus producing explicitly written potentials given as $V=V\bigl(z(x)\bigr)$ through an elementary function $z=z(x)$. All these cases are achieved by fixing the parameter $a$ to a particular value, hence, all these particular potentials are four-parametric. The mentioned two recently presented four-parametric ordinary hypergeometric potentials [9,10] are just such cases.

The potential we present is either a singular well (which behaves as the inverse square root in the vicinity of the origin and exponentially vanishes at infinity) or a smooth asymmetric step-barrier (with variable height, steepness, and asymmetry). The general solution of the Schrödinger equation for this potential is written through fundamental solutions each of which presents an irreducible linear combination of two ordinary hypergeometric functions $_2F_1$. The singular version of the potential describes a short-range interaction and for this reason supports only a finite number of bound states. We derive the exact equation for energy spectrum and estimate the number of bound states.

## 2. The potential

The potential is given parametrically as

$$V(z) = V_0 + \frac{V_1}{z}, \tag{1}$$

$$x(z) = x_0 + \sigma\big(a\ln(z-a) - \ln(z-1)\big), \tag{2}$$

where $a \neq 0,1$ and $x_0, \sigma, V_0, V_1$ are arbitrary (real or complex) constants. Rewriting the coordinate transformation as

$$\frac{(z-a)^a}{z-1} = e^{\frac{x-x_0}{\sigma}}, \tag{3}$$

it is seen that for real rational $a$ the transformation is rewritten as a polynomial equation for $z$, hence, in several cases it can be inverted. Since $a \neq 0,1$, the possible simplest case is when the polynomial equation is quadratic. This is achieved for $a = -1, 1/2, 2$. It is checked, however, that these three cases lead to four-parametric sub-potentials which are equivalent in the sense that each is derived from another by specifications of the involved parameters. For $a = -1$ the potential reads [9]:

$$V(x) = V_0 + \frac{V_1}{\sqrt{1 + e^{(x-x_0)/\sigma}}}. \tag{4}$$

The next are the cubic polynomial reductions of equation (3) which are achieved in six cases: $a = -2, -1/2, 1/3, 2/3, 3/2, 3$. It is again checked, however, that these choices produce only one independent potential. This is the four-parametric potential presented in [10]:

$$V = V_0 + \frac{V_1}{z}, \quad z = -1 + \frac{1}{\left(e^{x/(2\sigma)} + \sqrt{1+e^{x/\sigma}}\right)^{2/3}} + \left(e^{x/(2\sigma)} + \sqrt{1+e^{x/\sigma}}\right)^{2/3}, \tag{5}$$

where one should replace $x$ by $x - x_0$. Similar potentials in terms of elementary functions through quartic and quintic reductions of equation (3) are rather cumbersome; we omit those.

For arbitrary real $a \neq 0,1$, assuming $z \in (0,1)$ and shifting

$$x_0 \to x_0 - \sigma a \ln(-a) + i\pi\sigma, \tag{6}$$

the potential (1),(2) presents a singular well. In the vicinity of the origin it behaves as $x^{-1/2}$:

$$V\big|_{x\to 0} \sim \sqrt{\frac{(a-1)\sigma}{2a}} \frac{V_1}{\sqrt{x}}, \tag{7}$$

and exponentially approaches a constant, $V_0 + V_1$, at infinity:

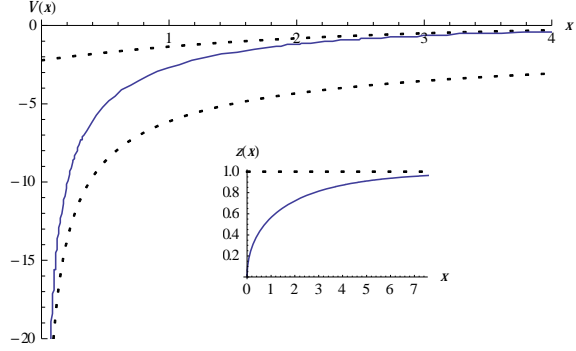

Fig.1. Potential (1), (2) for $a = -2$ and $(\sigma, x_0, V_0, V_1) = (2, 0, 5, -5)$.
The inset presents the coordinate transformation $z(x) \in (0,1)$ for $x \in (0, \infty)$.

$$V\big|_{x \to +\infty} \sim \left(\frac{a-1}{a}\right)^a V_1 e^{-x/\sigma}, \qquad (8)$$

The potential and the two asymptotes are shown in Fig. 1.

A potential of a different type is constructed if one allows the parameterization variable $z$ to vary within the interval $z \in (1, \infty)$ for $a < 1$ or within the interval $z \in (1, a)$ for $a > 1$. This time, shifting (compare with Eq. (6))

$$x_0 \to x_0 - \sigma a \ln(1-a), \qquad (9)$$

we get an asymmetric step-barrier the height of which depends on $V_0$ and $V_1$, while the asymmetry and steepness are controlled by the parameters $a$ and $\sigma$. The shape of the potential is shown in Fig. 2 for $a = -2$ and $a = 1.25$. We note that in the limit $\sigma \to 0$ the potential turns into the abrupt-step potential and that the sub-family of barriers generated by variation of $\sigma$ at constant $V_0$ and $V_1$ has a $\sigma$-independent fixed point located at $x = x_0$ (marked in Fig. 2 by filled circles).

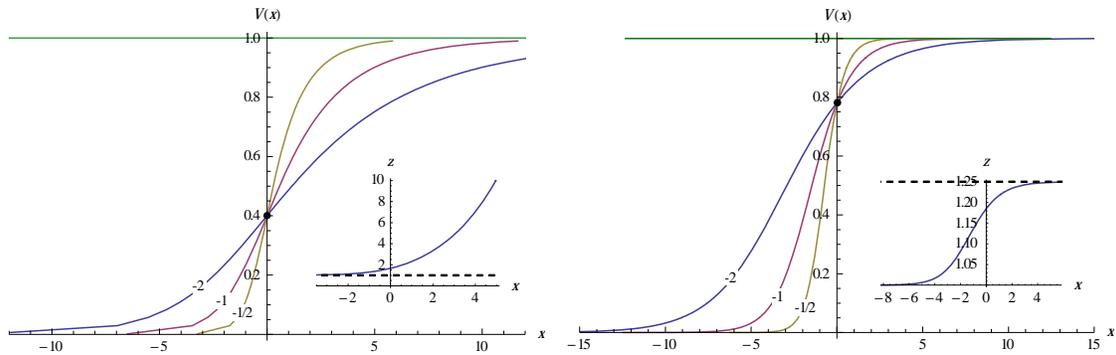

Fig.2. Potential (1),(2) for $a = -2$ and $(x_0, V_0, V_1) = (0, 1, -1)$ (left figure) and for $a = 1.25$ and $(x_0, V_0, V_1) = (0, 5, -5)$ (right figure); $\sigma = -2, -1, -1/2$. The fixed points are marked by filled circles. The insets present the coordinate transformation $z(x)$ for $\sigma = -1$.

## 2. Reduction to the general Heun equation

The solution of the one-dimensional Schrödinger equation for potential (1),(2):

$$\frac{d^2\psi}{dx^2} + \frac{2m}{\hbar^2}(E - V(x))\psi = 0, \tag{10}$$

is constructed via reduction to the general Heun equation [27-29]

$$\frac{d^2u}{dz^2} + \left(\frac{\gamma}{z-a_1} + \frac{\delta}{z-a_2} + \frac{\varepsilon}{z-a_3}\right)\frac{du}{dz} + \frac{\alpha\beta z - q}{(z-a_1)(z-a_2)(z-a_3)}u = 0. \tag{11}$$

The details of the technique are presented in [11] and [25]. It has been shown that the energy-independent general-Heun potentials, which are proportional to an arbitrary variable parameter and have shapes which are independent of that parameter, are constructed by the coordinate transformation $z = z(x)$ of the Manning form [30] given as

$$\frac{dz}{dx} = (z-a_1)^{m_1}(z-a_2)^{m_2}(z-a_3)^{m_3}/\sigma, \tag{12}$$

where $m_{1,2,3}$ are integers or half-integers and $\sigma$ is an arbitrary scaling constant. As it is seen, the coordinate transformation is solely defined by the singularities $a_{1,2,3}$ of the general Heun equation. The canonical form of the Heun equation assumes two of the three finite singularities at $0$ and $1$, and the third one at a point $a$, so that $a_{1,2,3} = (0,1,a)$ [27-29]. However, it may be convenient for practical purposes to apply a different specification of the singularities, so for the moment we keep the parameters $a_{1,2,3}$ unspecified.

The coordinate transformation is followed by the change of the dependent variable

$$\psi = (z-a_1)^{\alpha_1}(z-a_2)^{\alpha_2}(z-a_3)^{\alpha_3}u(z) \tag{13}$$

and application of the ansatz

$$V(z) = \frac{v_0 + v_1 z + v_2 z^2 + v_3 z^3 + v_4 z^4}{(z-a_1)^2(z-a_2)^2(z-a_3)^2}\left(\frac{dz}{dx}\right)^2, \quad v_{0,1,2,3,4} = \text{const}. \tag{14}$$

The form of this ansatz and the permissible sets of the parameters $m_{1,2,3}$ are revealed through the analysis of the behavior of the solution in the vicinity of the finite singularities of the general Heun equation [11]. This is a crucial point which warrants that all the parameters involved in the resultant potentials can be varied independently.

It has been shown that there exist in total thirty-five permissible choices for the coordinate transformation each being defined by a triad $(m_1, m_2, m_3)$ satisfying the inequalities $-1 \leq m_{1,2,3} \leq 1$ and $1 \leq m_1 + m_2 + m_3 \leq 3$ [25]. However, because of the symmetry of the general Heun equation with respect to the transpositions of its singularities, only eleven of the resultant potentials are independent [25]. The potential (1),(2) belongs to the fifth independent family with $m_{1,2,3} = (1,1,-1)$ for which from equation (14) we have

$$V(z) = \frac{V_4 + V_3 z + V_2 z^2 + V_1 z^3 + V_0 z^4}{(z-a_3)^4} \tag{15}$$

with arbitrary $V_{0,1,2,3,4} = \text{const}$, and, from equation (12),

$$\frac{x-x_0}{\sigma} = \frac{a_1 - a_3}{a_1 - a_2} \ln(z-a_1) + \frac{a_3 - a_2}{a_1 - a_2} \ln(z-a_2). \tag{16}$$

It is now convenient to have a potential which does not explicitly involve the singularities. Hence, we put $a_3 = 0$ and apply the specification $a_{1,2,3} = (a,1,0)$ to get the potential

$$V(z) = V_0 + \frac{V_1}{z} + \frac{V_2}{z^2} + \frac{V_3}{z^3} + \frac{V_4}{z^4} \tag{17}$$

with

$$\frac{x-x_0}{\sigma/(a-1)} = a \ln(z-a) - \ln(z-1). \tag{18}$$

The solution of the Schrödinger equation (10) for this potential is written in terms of the general Heun function $H_G$ as

$$\psi = (z-a)^{\alpha_1} z^{\alpha_2} (z-1)^{\alpha_3} H_G(a_1, a_2, a_3; q; \alpha, \beta, \gamma, \delta, \varepsilon; z), \tag{19}$$

where the involved parameters $\alpha, \beta, \gamma, \delta, \varepsilon$ and $q$ are given through the parameters $V_{0,1,2,3,4}$ of the potential (17) and the exponents $\alpha_{1,2,3}$ of the pre-factor by the equations [25]

$$(\gamma, \delta, \varepsilon) = (1+2\alpha_1, 1+2\alpha_2, -1+2\alpha_3), \tag{20}$$

$$1+\alpha+\beta = \gamma+\delta+\varepsilon, \quad \alpha\beta = (\alpha_1+\alpha_2+\alpha_3)^2 + 2m\sigma^2(E-V_0)/\hbar^2, \tag{21}$$

$$q = \frac{2m\sigma^2}{\hbar^2}(V_1 - (1+a)(E-V_0)) + \tag{22}$$

$$(-\alpha_2^2 + (-1+\alpha_1+\alpha_3)(\alpha_1+\alpha_3)) + a(-\alpha_1^2 + (-1+\alpha_2+\alpha_3)(\alpha_2+\alpha_3));$$

the exponents $\alpha_{1,2,3}$ of the pre-factor being defined by the equations

$$\alpha_1^2 = \frac{2m\sigma^2}{a^2(a-1)^2\hbar^2}(V_4 + aV_3 + a^2V_2 + a^3V_1 + a^4(V_0 - E)), \tag{23}$$

$$\alpha_2^2 = -\frac{2m\sigma^2}{(a-1)^2\hbar^2}(E-V_0-V_1-V_2-V_3-V_4), \tag{24}$$

$$\alpha_3(\alpha_3-2) = \frac{2m\sigma^2 V_4}{a^2\hbar^2}. \tag{25}$$

## 3. The solution of the Schrödinger equation in terms of the Gauss functions

Having determined the parameters of the Heun equation, the next step is to examine the cases when the general Heun function $H_G$ is written in terms of the Gauss hypergeometric functions $_2F_1$. An observation here is that the direct one-term Heun-to-hypergeometric reductions discussed by many authors (see, e.g., [27-28, 31-34]) are achieved by such restrictions, imposed on the involved parameters (three or more conditions), which are either not satisfied by the Heun potentials or produce very restrictive potentials. It is checked that the less restrictive reductions reproduce the classical Eckart and Pöschl-Teller potentials, while the other reductions result in conditionally integrable potentials.

More advanced are the finite-sum solutions achieved by termination of the series expansions of the general Heun function in terms of the hypergeometric functions [35-39]. For such reductions, only two restrictions are imposed on the involved parameters and, notably, these restrictions are such that in many cases they are satisfied. The solutions for the above-mentioned four-parametric sub-potentials [9,10] have been constructed right in this way. Other examples achieved by termination of the hypergeometric series expansions of the functions of the Heun class include the recently reported inverse square root [4], Lambert-W step [5] and Lambert-W singular [6] potentials.

The series expansions of the general Heun function in terms of the Gauss ordinary hypergeometric functions are governed by three-term recurrence relations for the coefficients of the successive terms of the expansion. A useful particular expansion in terms of the functions of the form $_2F_1(\alpha,\beta;\gamma_0-n;z)$ which leads to simpler coefficients of the recurrence relation is presented in [25,39]. If the expansion functions are assumed irreducible to simpler functions, the termination of this series occurs if $\varepsilon=-N$, $n=0,1,2,...$, and a $(N+1)$th degree polynomial equation for the accessory parameter $q$ is satisfied. For $\varepsilon=0$ the latter equation is $q=a\alpha\beta$, which corresponds to the trivial direct reduction of the general Heun equation to the Gauss hypergeometric equation. This case reproduces the classical Eckart and Pöschl-Teller potentials [25]. For the first nontrivial case $\varepsilon=-1$ the termination condition for singularities $a_{1,2,3}=(a,1,0)$ takes a particularly simple form:

$$q^2+q(\gamma-1+a(\delta-1))+a\alpha\beta=0. \qquad (26)$$

The solution of the Heun equation for a root of this equation is written as [39]

$$u = {}_2F_1\left(\alpha,\beta;\gamma;\frac{a-z}{a-1}\right)+\frac{\gamma-1}{q+a(\delta-1)}\cdot{}_2F_1\left(\alpha,\beta;\gamma-1;\frac{a-z}{a-1}\right), \qquad (27)$$

This solution has an alternative representation through Clausen's generalized hypergeometric function $_3F_2$ [40,41].

Consider if the termination condition (26) for $\varepsilon = -1$ is satisfied for the parameters given by equations (20)-(25). From the last equation (20) we get that for $\varepsilon = -1$ holds $\alpha_3 = 0$. It then follows from Eq. (25) that $V_4 = 0$. With this, equation Eq. (26) is reduced to

$$V_2 + V_3\left(\frac{1+a}{a} - \frac{2m\sigma^2}{a^2\hbar^2}V_3\right) = 0. \tag{28}$$

This equation generally defines a conditionally integrable potential in that the potential parameters $V_2$ and $V_3$ are not varied independently. Alternatively, if the potential parameters are assumed independent, the equation is satisfied only if $V_2 = V_3 = 0$. Thus, we put $V_{2,3,4} = 0$ and potential (17) is reduced to that given by equation (1). Furthermore, since $\sigma$ is arbitrary, in order for equation (18) to exactly reproduce the coordinate transformation (2), we replace $\sigma/(1-a) \to \sigma$.

With this, the solution of the Schrödinger equation (10) for potential (1) is written as

$$\psi = (z-a)^{\alpha_1}(z-1)^{\alpha_2}\left(\,_2F_1\left(\alpha,\beta;\gamma;\frac{a-z}{a-1}\right) + \frac{2\alpha_1}{a\alpha_2 - \alpha_1} \cdot \,_2F_1\left(\alpha,\beta;\gamma-1;\frac{a-z}{a-1}\right)\right) \tag{29}$$

with
$$(\alpha,\beta,\gamma) = (\alpha_1 + \alpha_2 + \alpha_0, \alpha_1 + \alpha_2 - \alpha_0, 1 + 2\alpha_1), \tag{30}$$

$$\alpha_{0,1,2} = \left(\pm\sqrt{\frac{2m\sigma^2(a-1)^2}{\hbar^2}(V_0 - E)},\; \pm\sqrt{\frac{2m\sigma^2 a^2}{\hbar^2}\left(V_0 - E + \frac{V_1}{a}\right)},\; \pm\sqrt{\frac{2m\sigma^2}{\hbar^2}(V_0 - E + V_1)}\right). \tag{31}$$

This solution applies for any real or complex set of the involved parameters. Furthermore, we note that any combination for the signs of $\alpha_{1,2}$ is applicable. Hence, by choosing different combinations, one can construct different independent fundamental solutions. Thus, this solution supports the general solution of the Schrödinger equation.

A final remark is that using the contiguous functions relations for the hypergeometric functions [43] one can replace the second hypergeometric function in Eq. (29) by a linear combination of the first hypergeometric function and its derivative. In this way we arrive at the following representation of the general solution of the Schrödinger equation:

$$\psi = (z-a)^{\alpha_1}(z-1)^{\alpha_2}\left(F + \frac{z-a}{\alpha_1 + a\alpha_2}\frac{dF}{dz}\right), \tag{32}$$

where
$$F = c_1 \cdot \,_2F_1\left(\alpha,\beta;\gamma;\frac{a-z}{a-1}\right) + c_2 \cdot \,_2F_1\left(\alpha,\beta;1+\alpha+\beta-\gamma;\frac{z-1}{a-1}\right). \tag{33}$$

## 4. Bound states

Consider the bound states supported by the singular version of potential (1),(2), achieved by shifting $x_0 \to x_0 - \sigma a \ln(-a) + i\pi\sigma$ in Eq. (2). Since the potential vanishes at infinity exponentially, it is understood that this is a short-range potential. The integral of the function $xV(x)$ over the semi-axis $x \in (0,+\infty)$ is finite, hence, according to the general criterion [43-47], the potential supports only a finite number of bound states. These states are derived by demanding the wave function to vanish both at infinity and in the origin (see the discussion in [48]). We recall that for this potential the coordinate transformation maps the interval $x \in (0,+\infty)$ onto the interval $z \in (0,1)$. Thus, we demand $\psi(z=0) = \psi(z=1) = 0$.

The condition $V(+\infty) = 0$ assumes $V_0 + V_1 = 0$, hence, $\alpha_2$ is real for negative energy. Choosing, for definiteness, the plus signs in Eq. (31), we have $\alpha_2 > 0$. Then, examining the equation $\psi(z=1) = 0$, we find that

$$\psi\big|_{z \to 1} \sim c_1 A_1 (1-z)^{-\alpha_2} + c_2 A_2 (1-z)^{\alpha_2} \tag{34}$$

with some constants $A_{1,2}$. Since for positive $\alpha_2$ the first term diverges, we conclude $c_1 = 0$. The condition $\psi(z=0) = 0$ then gives the following exact equation for the spectrum:

$$S(E) \equiv 1 + \frac{\alpha_1 + a\alpha_2}{2(1-a)\alpha_2} \frac{{}_2F_1\left(\alpha+1, \beta+1; 1+2\alpha_2; \frac{1}{1-a}\right)}{{}_2F_1\left(\alpha, \beta; 2\alpha_2; \frac{1}{1-a}\right)} = 0. \tag{35}$$

The graphical representation of this equation is shown in Fig. 3. The function $S(E)$ has a finite number of zeros. For the parameters $m, \hbar, V_0, \sigma, a = 1, 1, 5, 2, -2$ applied in the figure there are just three bound states.

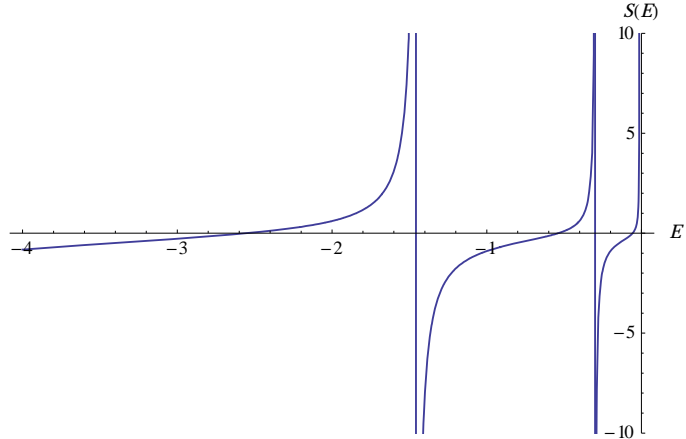

Fig.3. Graphical representation of the spectrum equation (35) for $m, \hbar, V_0, \sigma, a = 1, 1, 5, 2, -2$.

According to the general theory, the number of bound states is equal to the number of zeros (not counting $x=0$) of the zero-energy solution, which vanishes at the origin [43-47]. We note that for $E=0$ the lower parameter of the second hypergeometric function in Eq. (33) vanishes: $1+\alpha+\beta-\gamma=0$. Hence, a different second independent solution should be applied. This solution is constructed by using the first hypergeometric function with $\alpha_1$ everywhere replaced by $-\alpha_1$. The result is rather cumbersome. It is more conveniently written in terms of the Clausen functions as

$$\psi_{E=0} = c_1 (z-a)^{\alpha_1} {}_3F_2\left(-\sqrt{\frac{a-1}{a}}\alpha_1+\alpha_1, \sqrt{\frac{a-1}{a}}\alpha_1+\alpha_1, 1+\alpha_1; \alpha_1, 1+2\alpha_1; \frac{a-z}{a-1}\right) + $$
$$c_2 (z-a)^{-\alpha_1} {}_3F_2\left(-\sqrt{\frac{a-1}{a}}\alpha_1-\alpha_1, \sqrt{\frac{a-1}{a}}\alpha_1-\alpha_1, 1-\frac{\alpha_1}{a}; -\frac{\alpha_1}{a}, 1; \frac{z-1}{a-1}\right), \quad (36)$$

where $\alpha_1 = \sqrt{2a(a-1)m\sigma^2 V_0/\hbar^2}$ and the relation between $c_1$ and $c_2$ is readily derived from the condition $\psi_{E=0}(0)=0$. This solution is shown in Fig. 4. It is seen that for parameters $m,\hbar,V_0,\sigma,a = 1,1,5,2,-2$ used in Fig. 3 the number of zeros (excluded the origin) is indeed 3.

For practical purposes, it is useful to have an estimate for the number of bound states. The absolute upper limit for this number is given by the integral [43,44]

$$I_B = \int_0^\infty r\left|V(x\to r\hbar/\sqrt{2m})\right| dr = (1-a)\left(Li_2\left(\frac{1}{1-a}\right) + 2a\coth^{-1}(1-2a)^2\right)\frac{2m\sigma^2 V_0}{\hbar^2}. \quad (37)$$

where $Li_2$ is Jonquière's polylogarithm function of order 2 [49,50]. Though of general importance, however, in many cases this is a rather overestimating limit. Indeed, for the parameters applied in Fig. 3 it gives $n \leq I_B \approx 24$.

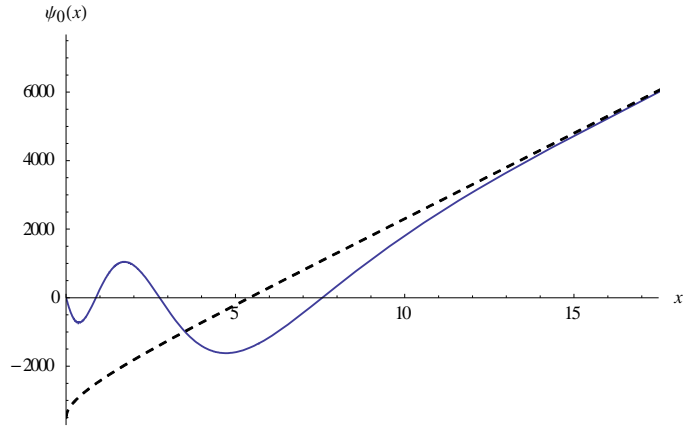

Fig.4. The zero-energy solution for $m,\hbar,V_0,\sigma,a = 1,1,5,2,-2$. The dashed line shows the logarithmic asymptote at infinity: $\psi_0\big|_{x\to\infty} \sim A+B\ln(1-z)$.

More stringent are the estimates by Calogero [45] and Chadan [46] which are specialized for everywhere monotonically non-decreasing attractive central potentials. Calogero's estimate reads $n \leq I_C$ with [45]

$$I_C = \frac{2/\pi}{\hbar/\sqrt{2m}} \int_0^\infty \sqrt{-V(x)}\,dx = \left(1+\left(\sqrt{1-a}-\sqrt{-a}\right)^2\right)\sqrt{\frac{2m\sigma^2 V_0}{\hbar^2}}, \qquad (38)$$

We note that $I_C \approx \sqrt{2I_B}$. The result by Chadan further tunes the upper limit for the number of bound states to the half of that by Calogero, that is $n \leq I_C/2$ [46]. For the parameters applied in Fig. 3 this gives $n \leq 3.48$, which is, indeed, an accurate estimate. The dependence of the function $n_c = I_C/2$ on the parameter $a$ for $a \in (-\infty, 0) \cup (1, +\infty)$ is shown in Fig. 5. It is seen that more bound states are available for $a$ close to zero. The maximum number achieved in the limit $a \to 0$ is $\sqrt{2m\sigma^2 V_0/\hbar^2}$, hence, for sufficiently small $V_0$ or $\sigma$ such that $2m\sigma^2 V_0 < \hbar^2$, bound states are not possible at all.

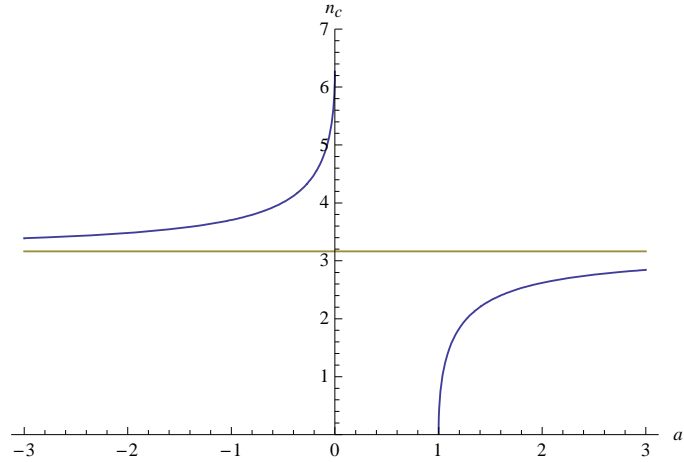

Fig.5. The dependence of Chadan's estimate $n_c = I_C/2$ for the number of bound states on the parameter $a$ ($m, \hbar, V_0, \sigma = 1, 1, 5, 2$).

## 5. Discussion

Thus, we have presented the third five-parametric quantum-mechanical potential for which the solution of the Schrödinger equation is written in terms of the Gauss ordinary hypergeometric functions. The potential involves five (generally complex) parameters which are varied independently. Depending on the particular specifications of these parameters, the potential suggests two different appearances. In one version we have a smooth step-barrier with variable height, steepness and asymmetry, while in the other version this is a singular potential-well which behaves as the inverse square root in the vicinity of the origin and exponentially vanishes at infinity.

The potential is in general given parametrically; however, in several cases the involved coordinate transformation allows inversion thus leading to particular potentials which are explicitly written in terms of elementary functions. These reductions are achieved by particular specifications of a parameter standing for the third finite singularity of the general Heun equation. The resultant sub-potentials all are four-parametric (see, e.g., [9,10]). These particular cases are defined by coordinate transformations which are roots of polynomial equations. It turns out that different polynomial equations of the same degree produce the same potential (with altered parameters). The reason for this is well understood in the case of quadratic equations. In that case the third singularity of the general Heun equation, to which the Schrödinger equation is reduced, is specified as $a = -1, 1/2$ or $2$. We then note that the form-preserving transformations of the independent variable map the four singularities of the Heun equation, $z = 0, 1, a, \infty$, onto the points $z = 0, 1, a_1, \infty$ with $a_1$ adopting one of the six possible values $a, 1/a, 1-a, 1/(1-a), a/(1-a), (a-1)/a$ [27-29]. It is seen that the triad $(-1, 1/2, 2)$ is a specific set which remains invariant at form-preserving transformations of the independent variable.

The potential belongs to the general Heun family $m_{1,2,3} = (1, 1, -1)$. This family allows several conditionally integrable reductions too [25]. A peculiarity of the exactly integrable potential that we have presented here is that the location of a finite singularity of the general Heun equation is not fixed to a particular point of the complex $z$-plane but serves as a variable potential-parameter. In the step-barrier version of the potential, this parameter stands for the asymmetry of the potential.

The solution of the Schrödinger equation for the potential we have presented is constructed via termination of a series expansion of the general Heun equation in terms of the Gauss ordinary hypergeometric functions. The general solution of the problem is composed of fundamental solutions each of which is an *irreducible* combination of two hypergeometric functions. Several other potentials allowing solutions of this type have been reported recently [4-6,9-10,23-25]. Further cases involve the solutions for super-symmetric partner potentials much discussed in the past [51-53] and for several non-analytic potentials discussed recently [54-56]. One should distinguish these solutions from the case of reducible hypergeometric functions [57-61] when the solutions eventually reduce to quasi-polynomials, e.g., discussed in the context of quasi-exactly solvability [59-61]. We note that owing to the contiguous functions relations [42], the two-term structure of the solution is a general property of all finite-sum hypergeometric reductions of the general Heun functions achieved via termination of series solutions. It is checked that in our case the linear combination of the involved Gauss functions is expressed through a single generalized hypergeometric function $_3F_2$ [40,41].

We have presented the explicit solution of the problem and discussed the bound states supported by the singular version of the potential. We have derived the exact equation for the energy spectrum and estimated the number of bound states. The exact number of bound states is given by the number of zeros of the zero-energy solution which we have also presented.

**Acknowledgments**

This research has been conducted within the scope of the International Associated Laboratory IRMAS (CNRS-France & SCS-Armenia). The work has been supported by the Armenian State Committee of Science (SCS Grant No. 15T-1C323), Armenian National Science and Education Fund (ANSEF Grant No. PS-4558) and the project "Leading Russian Research Universities" (Grant No. FTI_24_2016 of the Tomsk Polytechnic University). T.A. Ishkhanyan acknowledges the support from SPIE through a 2017 Optics and Photonics Education Scholarship and thanks the French Embassy in Armenia for a doctoral grant.

Reports **642**, 1-71 (2016).